\documentclass[useAMS,usenatbib]{mn2e/mn2e}
\usepackage[dvipsnames,usenames,table]{xcolor}
\usepackage{graphicx}
\usepackage{txfonts}
\usepackage{hyperref}
\usepackage{amssymb}
\def\aj{AJ}%
\def\apj{ApJ}%
\def\apjl{ApJ}%
\def\apjs{ApJS}%
\def\aap{A\&A}%
\def\aapr{A\&A~Rev.}%
\def\mnras{MNRAS}%
\def\pasp{PASP}%
\newcommand {\reac}[6] {$\rm\,{}^{#2}\kern-0.8pt{#1}\,({#3}\,,{#4})\,{}^{#6}\kern-0.8pt{#5}\,$}

\newcommand{\diff}{\mbox{${\rm d}$}}

\newcommand{\feh}{\mbox{\rm [{\rm Fe}/{\rm H}]}}

\newcommand{\Msun}{\mbox{$\mathrm{M}_{\odot}$}}

\newcommand{\Teff}{\mbox{$T_{\rm eff}$}}
\newcommand{\geff}{\mbox{$g$}}
\newcommand{\geffpol}{\mbox{$g_\mathrm{pol}$}}

\newcommand{\Rpol}{\mbox{$R_\mathrm{pol}$}}

\newcommand{\omegai}{\mbox{$\omega_\mathrm{i}$}}
\newcommand{\logg}{\mbox{$\log g$}}

\newcommand{\beq}{\begin{equation}}
\newcommand{\eeq}{\end{equation}}
\newcommand{\beqa}{\begin{eqnarray}}
\newcommand{\eeqa}{\end{eqnarray}}

\title[Photometry of fast rotators]{On the photometric signature of fast rotators}

\author[Girardi et al.]{
L\'eo Girardi$^1$, 
Guglielmo Costa$^2$, 
Yang Chen$^3$, 
Paul Goudfrooij$^4$, 
Alessandro Bressan$^2$, 
\newauthor
Paola Marigo$^3$, 
and Andrea Bellini$^4$
\\
$^1$ Osservatorio Astronomico di Padova -- INAF, Vicolo dell'Osservatorio 5, I-35122 Padova, Italy \\
$^2$ SISSA, via Bonomea 365, I-34136 Trieste, Italy \\
$^3$ Dipartimento di Fisica e Astronomia Galileo Galilei, Universit\`a di Padova, Vicolo dell'Osservatorio 3, I-35122 Padova, Italy \\
$^4$ Space Telescope Science Institute, 3700 San Martin Drive, Baltimore, MD 21218, USA \\
}

\begin{document}
\date{Received .... / Accepted .....}

\pagerange{\pageref{firstpage}--\pageref{lastpage}} \pubyear{2016}

\maketitle

\label{firstpage}

\begin{abstract}
Rapidly rotating stars have been recently recognized as having a major role in the interpretation of colour-magnitude diagrams of young and intermediate-age star clusters in the Magellanic Clouds and in the Milky Way. In this work, we evaluate the distinctive spectra and distributions in colour-colour space that follow from the presence of a substantial range in effective temperatures across the surface of fast rotators. The calculations are inserted in a formalism similar to the one usually adopted for non-rotating stars, which allows us to derive tables of bolometric corrections as a function not only of a reference effective temperature, surface gravity and metallicity, but also of the rotational speed with respect to the break-up value, $\omega$, and the inclination angle, $i$.  We find that only very fast rotators ($\omega>0.95$) observed nearly equator-on ($i>45^\circ$) present sizable deviations from the colour-colour relations of non-rotating stars. In light of these results, we discuss the photometry of the $\sim$\,200-Myr-old cluster NGC~1866 and its split main sequence, which has been attributed to the simultaneous presence of slow and fast rotators. The small dispersion of its stars in colour-colour diagrams allow us to conclude that fast rotators in this cluster either have rotational velocities $\omega<0.95$, or are all observed nearly pole-on. Such geometric colour-colour effects, although small, might be potentially detectable in the huge, high-quality photometric samples in the post-Gaia era, in addition to the evolutionary effects caused by rotation-induced mixing. 
\end{abstract}

\begin{keywords}
  stars: general
\end{keywords}

%%%%%%%%%%%%%%%%%%%%%%%%%%%%%%%%%%%%%%%%%%%%%%%%%%%%%%%%%%%%%%%%%%%%%%%
\section{Introduction}
\label{intro}

The existence of rapidly rotating stars has been known for long, especially in the form of Be stars and among field A-type stars in the Milky Way \citep[][and references therein]{royer09, vanbelle12}. However, it was only recently recognized that they could be playing a very important role in determining the photometric properties of other commonly-studied stars, namely those born in populous star clusters in the Magellanic Clouds and in Milky Way open clusters. Indeed, fast rotators have been identified spectroscopically in both the Large Magellanic Cloud clusters NGC~1866 and NGC~1818 \citep[][respectively]{dupree17,marino18lmc} and in the Galactic open cluster M~11 \citep{marino18gal}, and seem to be linked to the presence of extended main sequence turn-offs \citep[see e.g.][]{brandt15b} and/or split main sequences \citep[e.g.][]{milone16,milone17} in these objects. Since the same clusters have often been used to calibrate non-rotating stellar models, the neglect of rotation might have had important consequences. For instance, it might have caused systematic errors in the age estimates of well-studied clusters \citep{brandt15c, gossage18}, and an overestimation of the amount of convective core overshooting needed to reproduce the luminosity of post-main sequence stars \citep{costa19}. 

With this in mind, it is very important that we expand the previous sets of stellar models by including the effects of rotation, in a way suitable to the study of such clusters, as well as to the multitude of field stars which likely started their nuclear-burning lives as fast rotators. In the present work, we make a step in this direction, by evaluating the distinctive spectra and distributions in colour-colour space that follow from the presence of a substantial range in effective temperatures across the surface of fast rotators. We recall that similar computations have been performed a few times in the past literature. While many authors have concentrated their efforts on detailed calculations of changes in line profiles and equivalent widths in order to inform spectroscopic studies \citep[e.g.][]{slettebak80, fremat05}, the works by \citet{maeder72}, \citet{collins77}, \citet{perezhernandez99} and \citet{lovekin06} are especially insightful about the effects of rotation on the photometry.

Our calculations are intended to complement the libraries of evolutionary models for rotating stars based on the PAdova-tRieste Stellar Evolutionary Code \citep[PARSEC;][]{bressan12} code, which we started to present in \citet{costa19}, and will further detail in forthcoming papers. As a library of models intended for stellar population studies, they cover the widest possible range of parameters -- including the mean effective temperature and surface gravity, metallicity, angular velocity and inclination -- and photometric systems as well. As will be shown later, these new computations are necessary for a proper evaluation of the effects -- or signatures -- of rotation in the present databases of very accurate HST multi-band photometry of Magellanic Cloud clusters. 

Our input data and methods for the spectral computations are described in Sect.~\ref{data}, which also describes the formalism for the inclusion of such tables into isochrones and stellar population codes. Some properties of the derived spectra and colour-colour relations are illustrated in Sect.~\ref{results}, together with a couple of practical applications/considerations which are independent of stellar evolutionary models. Sect.~\ref{conclu} presents a few final comments and conclusions.

%%%%%%%%%%%%%%%%%%%%%%%%%%%%%%%%%%%%%%%%%%%%%%%%%%%%%%%%%%%%%%%%%%%%%%%
\section{Computing the spectra}
\label{data}

%%%%%%%%%%%%%%%%%%%%%%%%%%%%%%%%%%%%%%%%%%%%%%%%%%%%%%%%%%%%%%%%%%%%%%%
\subsection{The special case of non-rotating stars}
\label{sec_norot}

Before we deal with rotating stars, it is convenient to recall the formalism and assumptions involved in the computation of synthetic spectra of non-rotating, perfectly spherical stars.

The energy a stellar surface element of area $\diff A$, observed at an inclination angle $\theta$, emits per unit time $\diff t$' and wavelength $\diff\lambda$, over an element of solid angle $\diff\omega'$, is
\begin{equation}
    \diff E_\lambda = I_\lambda \diff A \cos\theta \diff t \diff\lambda \diff\omega'
\end{equation}
where $I_\lambda$ is the specific intensity. In a plane-parallel atmosphere, $I_\lambda$ depends only on $\theta$, and integration over the all solid angles provides the astrophysical net flux
\begin{eqnarray}
    F_\lambda &=& \int_{4\pi} I_\lambda(cos\theta) 
        \cos\theta \diff\omega' 
        \nonumber \\ 
        &=& \int_0^{2\pi}\int_0^{\pi/2} I_\lambda(\cos\theta) 
        \cos\theta \sin \theta \diff\theta \diff\phi
        \label{eq:deffnu}
\end{eqnarray}
where the final integral was limited to the interval $[0,\pi/2]$, because there is no incoming radiation from outside the star. The equation can be simplified further, with the definition of $\mu=\cos\theta$:
\begin{eqnarray}
    F_\lambda &=& 2\pi \int_0^1 I_\lambda(\mu) \mu \diff\mu  \,.
        \label{eq:specialcase}
\end{eqnarray}
This latter equation is actually used to compute $F_\lambda$ once $I_\lambda$ is given for several values of $\mu$, in a plane-parallel atmosphere calculation. $F_\lambda$ respects the Stefan-Boltzmann law
\begin{equation}
    \int_0^\infty F_\lambda \diff\lambda = F_\mathrm{bol} = \sigma\Teff^4  \,.
        \label{eq:stefan}
\end{equation}

The same symmetry that allows us to convert eq.~(\ref{eq:deffnu}) into eq.~(\ref{eq:specialcase}), leads to another important consequence: For a sphere of constant \Teff\ observed from a given line-of-sight, the $I_\lambda$ integrated over all surface elements in a half sphere, is proportional to the $I_\lambda$ of an element area averaged over all outgoing lines-of-sight, or, more precisely
\begin{eqnarray}
    \int_\mathrm{half-sphere} I_\lambda(\cos\theta) \cos\theta \diff A  =  R^2 F_\lambda
        \label{eq:sphereapprox}
\end{eqnarray}
which in practice allow us to avoid integrating the flux over the stellar surface. The monochromatic stellar luminosity becomes
\begin{equation}
    L_\lambda=4\pi R^2 F_\lambda ,
    \label{eq:lnu}
\end{equation}
and the flux observed at a distance $d$ of the star simply scales from the flux outgoing from a small piece of its surface:
\begin{equation}
    F_\lambda^\mathrm{observed} = \left(\frac{R}{d}\right)^2 F_\lambda^\mathrm{stellar\,surface}
    \label{eq:distance}
\end{equation}
where $F_\lambda^\mathrm{stellar\,surface}$ comes from eq.~(\ref{eq:specialcase}).

Equations (\ref{eq:specialcase}) to (\ref{eq:distance}) are at the basis of the synthetic photometry method, and the definition of bolometric corrections \citep[see][]{girardi02}.

%%%%%%%%%%%%%%%%%%%%%%%%%%%%%%%%%%%%%%%%%%%%%%%%%%%%%%%%%%%%%%%%%%%%%%%
\subsection{The case of rotating stars}

For rotating stars there is no spherical symmetry, and the simplification that leads from eq.~(\ref{eq:deffnu}) to eq.~(\ref{eq:sphereapprox}) does not apply. The outcoming flux has to be computed explicitly from integration over the visible surface of the star. 

To describe this surface, let us adopt the spherical coordinates $(r,\theta,\phi)$, with $\theta=0$ aligned with the rotation axis. For convenience, the stellar radius can be scaled to its polar value, i.e.\ $r=R(\theta)/\Rpol$. For a star of mass $M$ rotating at an angular velocity $\Omega$, in the Roche approximation, the isobar that defines the surface of radius $R$ as a function of the polar angle $\theta$ is given by 
\begin{equation}
    -\frac{GM}{R(\theta)}-\frac{1}{2}\Omega^2 R(\theta)^2 \sin^2\theta = -\frac{GM}{\Rpol} 
\end{equation}
which can be converted in \citep[see e.g.][for details]{maeder99, maederbook}
\begin{equation}
    \frac{4}{27}\omega^2 r^3 \sin^2(\theta) = r-1 
\end{equation}
and then solved numerically to give $r(\theta)$, for any value of the angular velocity with respect to its critical break-up value, $\omega = \Omega/\Omega_\mathrm{crit}$, where $\Omega_\mathrm{crit}=(2/3)^{3/2}\sqrt{GM/\Rpol^3}$. The local effective surface gravity, \geff, follows from computing the gravity and centrifugal forces at $(r,\theta)$, and can also be easily expressed in terms of its value at the stellar pole, $\geffpol=GM/\Rpol^2$. 
In addition, the surface radiative flux, hence the local $\Teff^4(\theta)$, scales with $\geff$ as:
\begin{equation}
   \Teff  \propto f(\theta,\omega) \, \geff^{1/4}
   \label{eq:espinosa}
\end{equation}
in which $f(\theta,\omega)$ is either equal to $1$ as in \citet{vonzeipel24}'s theorem, or the term coming from the solution of equation (24) in \citet{espinosalara11}. We adopt the latter formalism in this work. Unless all stellar quantities are specified (including $M$ and $\Rpol$), this equation tell us how \Teff\ varies across the surface, but not its absolute value. To fix the \Teff\ scale, we define the parameter $\Teff_0$, which is the \Teff\ value that a non-rotating star of the same $\Rpol$ should have to produce the same luminosity:
\begin{equation}
    \Teff_0 = 
    \left(
    \frac{1}{4\pi\Rpol^2} \int_\mathrm{surface} \Teff^4(\theta) \diff A 
    \right)^{1/4}\,.
\end{equation}
The right-hand side can be computed without actually specifying $\Rpol$, and its result is used to re-scale the entire $\Teff(\theta)$ relation up/down to a given $\Teff_0$.

\begin{figure}
\centering
\includegraphics[width=\columnwidth]{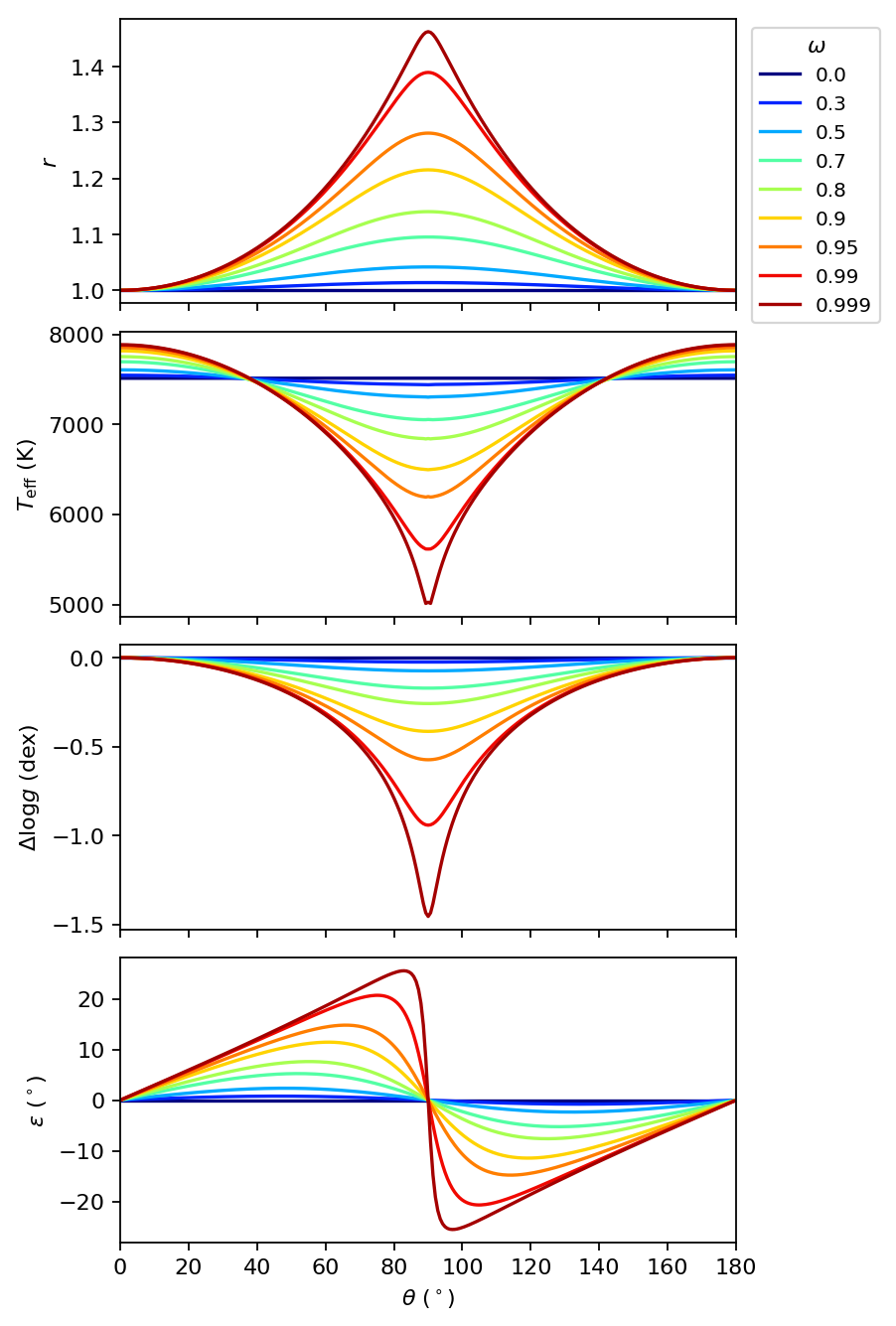}
\caption{The variation of stellar surface quantities with the angle $\theta$, for stars with $\Teff_0=7500$~K and several values of $\omega$. Panels from top to bottom show the stellar radius scaled to its polar value, the local effective temperature, the surface gravity with respect to its polar value, and the angle $\epsilon$ between the radius and the normal to the surface. }
\label{fig:theta_phi4000}
\end{figure}

These approximations and definitions suffice to give us all quantities of interest -- namely $r(\theta)$ $\Teff(\theta)$, and $\geff(\theta)$ -- as a function of our selected stellar input parameters, $\omega$, $\Teff_0$, and $\geffpol$. Another important variable is $\epsilon$, defined as the angle between the radial vector and the normal to the surface. For a ``flattened'' rotating star, $\epsilon$ is positive for $\theta<90^\circ$, then rapidly falls to null at the equator, then becomes negative for $90^\circ<\theta<180^\circ$. These quantities are illustrated in Fig.~\ref{fig:theta_phi4000}, for a few values of $\omega$ and for $\Teff_0=7500$~K. It can be noticed that the deviations from spherical symmetry (or, equivalently, from the polar values) are quite modest for all $\omega<0.5$, but increase dramatically as $\omega$ approaches 1.

With the coordinate system $(r,\theta,\phi)$ attached to the rotating star, we can now move the observer from the pole to the equator, defining a new angle $i$, which could be called ``viewing angle'', but in reality it is the ``inclination angle'' that a rotating star would have as seen on the sky ($i=0^\circ$ when observing from above the pole, $i=90^\circ$ when observing from the equator). In addition, let us define a quantity $F_\lambda^\mathrm{rot}$ that is similar to the $F_\lambda$ of eq.~(\ref{eq:distance}) -- in the sense that it could be used to compute the flux coming from distant rotating stars by means of:
\begin{equation}
    F_\lambda^\mathrm{rot, observed}(i) = \left(\frac{\Rpol}{d}\right)^2 F_\lambda^\mathrm{rot}(i)
    \label{eq:distancerot}
\end{equation}
Since the radius varies with $\theta$, in this later equation we have adopted the polar radius as the reference one. With this definition, we can write the analogous of eq.~(\ref{eq:sphereapprox}) for a rotating star:
%the $F_\lambda^\mathrm{rot}(i)$ can be computed from the integral of $I_\lambda$ over the surface of a rotating star with $\Rpol$:
\begin{eqnarray}
       F_\lambda^\mathrm{rot}(i) &=& \frac{1}{\Rpol^2}
       \int_\mathrm{surface} I_\lambda(\cos\xi)\cos\xi \diff A 
    \nonumber\\
       &=& \frac{1}{\Rpol^2}
       \int_0^{2\pi}\int_0^\pi  I_\lambda(\mu')\mu' (r^2/\cos\epsilon)\sin\theta \diff\theta \diff\phi      \,,
       \label{eq:integraloverrot}
\end{eqnarray}
where: 
\begin{itemize}
\item $\diff A=(r^2/\cos\epsilon)\sin\theta \, \diff\theta \diff\phi$ is the element of surface area for the coordinates $(\theta,\phi)$. Compared to the spherical case of eq.~(\ref{eq:sphereapprox}), it has two new multiplicative factors: $r^2$ takes into account the increase in area due to the variation in $r(\theta)$, while $(1/\cos\epsilon)$ takes into account the increase in area because the surface elements are inclined with respect to the radial vector.
\item $\xi$ is the angle between the normal to the surface $\vec{g}$ and the direction of the observer $\vec{i}$. The cosine of this angle, $\mu'$, is now used to select the proper $I_\lambda$ at every position, and to compute the projected area with respect to the observer (which is $\diff A\mu'$). Therefore, $\mu'$ replaces the former $\mu$ used for spherical stars.
\item As with the non-rotating case, the integration is performed only for surface elements in the visible part of the star, i.e.\ those with $\mu'$ between 1 and 0. 
\item Since the total stellar surface also scales with $\Rpol^2$, eq.~(\ref{eq:integraloverrot}) can be computed without actually specifying $\Rpol$. Moreover, since the $\Teff$ scale has been rescaled to produce a given $\Teff_0$ value, the total stellar luminosity relates simply to the $(\Rpol,\Teff_0)$ properties of our choice:
\begin{eqnarray}
  L &=& \int_0^\infty L_\lambda \diff \lambda 
  %\nonumber\\
  = \int_\mathrm{surface} \int_0^\infty F_\lambda \diff\lambda  \diff A 
  \nonumber\\
  &=& \int_\mathrm{surface} \sigma\Teff^4(\theta) \diff A 
  %\nonumber\\
  = 4\pi\Rpol^2 \sigma\Teff_0^4
\end{eqnarray}
\end{itemize}

In practice, we compute Eq.~(\ref{eq:integraloverrot}) numerically, by dividing the stellar surface into several thousands of pieces of size $\Delta\theta\times\Delta\phi$ not larger than a few square degrees each. All quantities are evaluated at every point of the stellar surface, with $\mu'$ values being re-evaluated for every inclination angle $i$. 

\begin{figure*}
\centering
\includegraphics[width=0.999\textwidth]{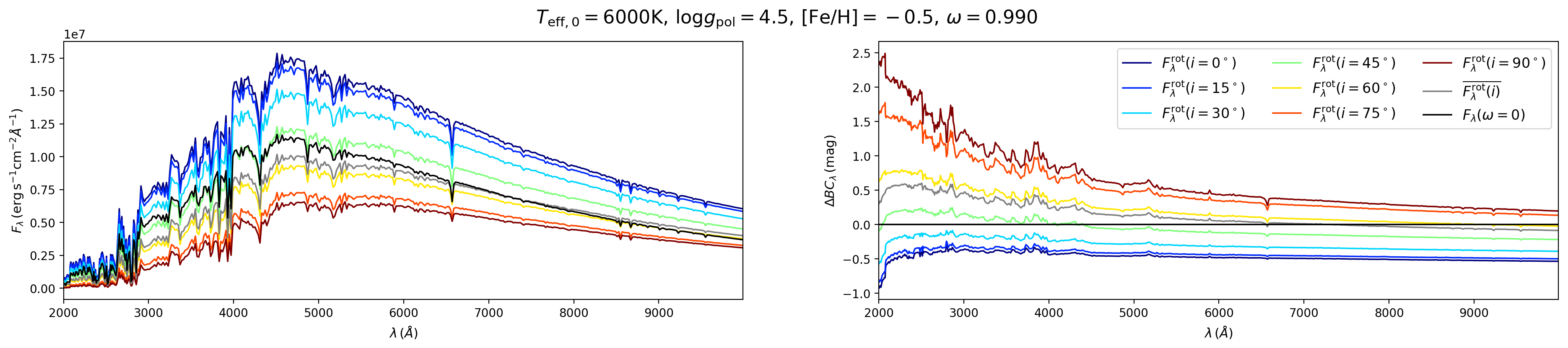}
\includegraphics[width=0.999\textwidth]{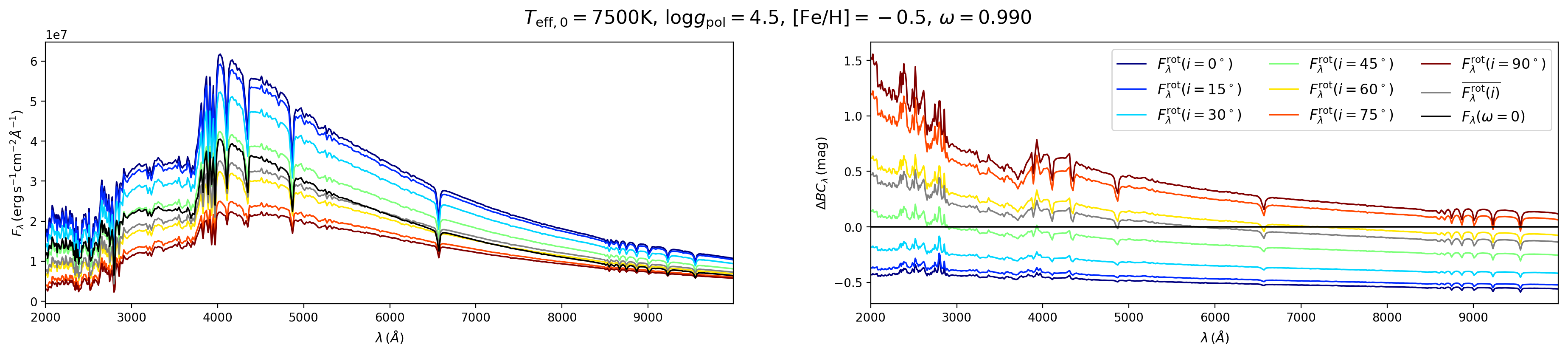}
\includegraphics[width=0.999\textwidth]{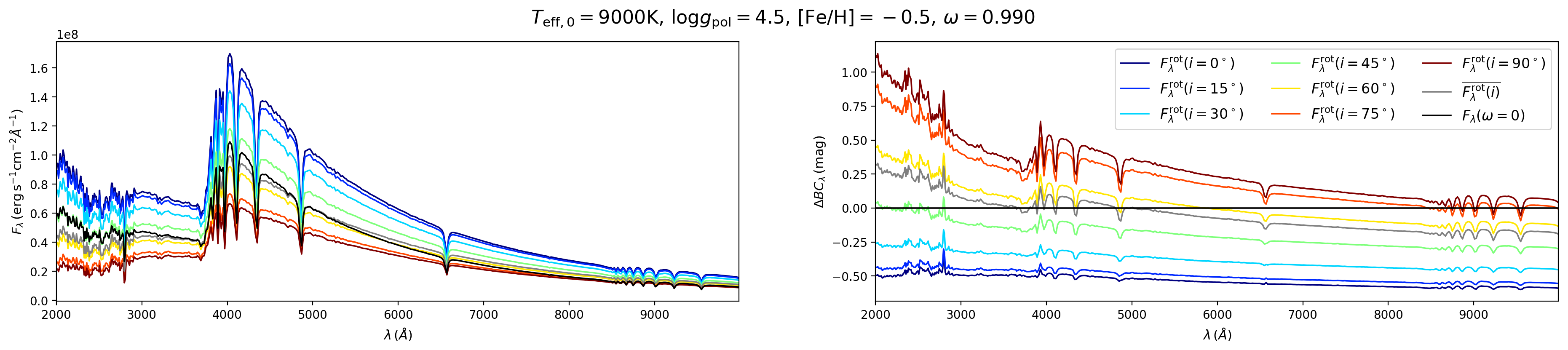}
\includegraphics[width=0.999\textwidth]{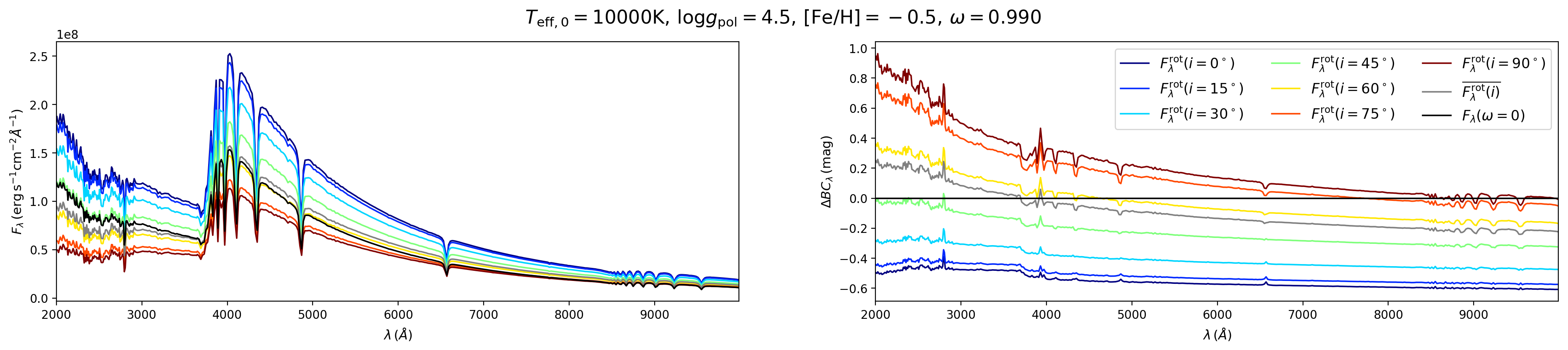}
\includegraphics[width=0.999\textwidth]{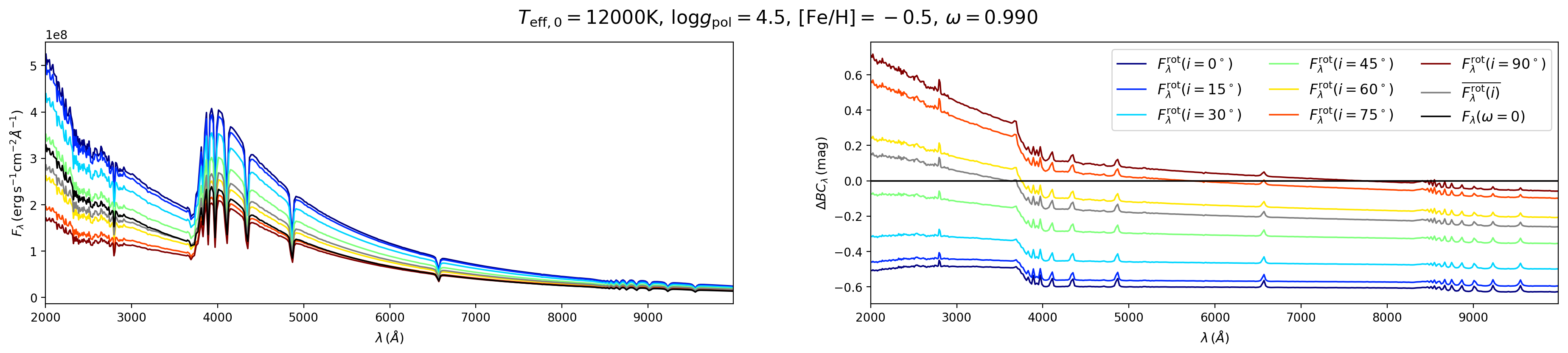}
\caption{Left panels: $F_\lambda^\mathrm{rot}(i)$ computed for a sequence of stars of increasing $\Teff_0$ (from 6000 to 12000~K, from top to bottom panels), and for a fixed $\omega=0.99$, $\logg_\mathrm{P}$, and $\mathrm{[Fe/H]}=-0.5$~dex. Coloured lines show the $F_\lambda$ observed from various angles from the pole ($i=0$, bluer) to the equator ($i=90^\circ$, redder). The gray line is the mean $F_\lambda$ averaged from all possible lines-of-sight, $\overline{F_\lambda^\mathrm{rot}(i)}$ (Sect.~\ref{sec:consistency}). Finally, the dark line is the $F_\lambda$ computed for the non-rotating star of same $\Teff_0$. Right panels: The same models but now plotting, in a magnitude scale, the relative flux of the rotating models compared to the reference non-rotating star -- that is, plotting $\Delta BC_\lambda=-2.5\log [F_\lambda(\omega=0.99)/F_\lambda(\omega=0)]$.
}
\label{fig:spectra}
\end{figure*}

Finally, several $F_\lambda^\mathrm{rot}(i)$ are shown in the left panels of Fig.~\ref{fig:spectra} for a rapidly-rotating star of $\omega=0.99$, and compared to similar quantities obtained for a spherical non-rotating star of same $\Teff_0$. It is evident that the star seen from the pole will be much brighter, and bluer, than when seen from the equator. The right panels allow us to appreciate these differences in a magnitude-like scale, by plotting the quantity 
\begin{equation}
    \Delta BC_\lambda=-2.5\log \frac{F_\lambda^\mathrm{rot}( \omega,i)}{F_\lambda(\omega=0)},
\end{equation}
where $F_\lambda(\omega=0)$ is the flux of the reference non-rotating star.

%%%%%%%%%%%%%%%%%%%%%%%%%%%%%%%%%%%%%%%%%%%%%%%%%%%%%%%%%%%%%%%%%%%%%%%
\subsection{Consistency and accuracy checks}
\label{sec:consistency}

Given the non-standard procedure adopted to compute these spectra, we perform a series of consistency and accuracy checks. First, we verify that our code recovers the $F_\lambda$ provided by \citet{castelli04}, for the case of $\omega=0$, for a wide range of \Teff\ (or, equivalently, $\Teff_0$) values, and for any value of $i$. 
Then, we modify the code so that we use the geometry from the rotating star, but the $I_\lambda(\Teff_0)$ everywhere. In this case, $F_\lambda^\mathrm{rot}(i)$ simply decreases with the projected area, as expected.

\begin{figure*}
\centering
\includegraphics[width=\textwidth]{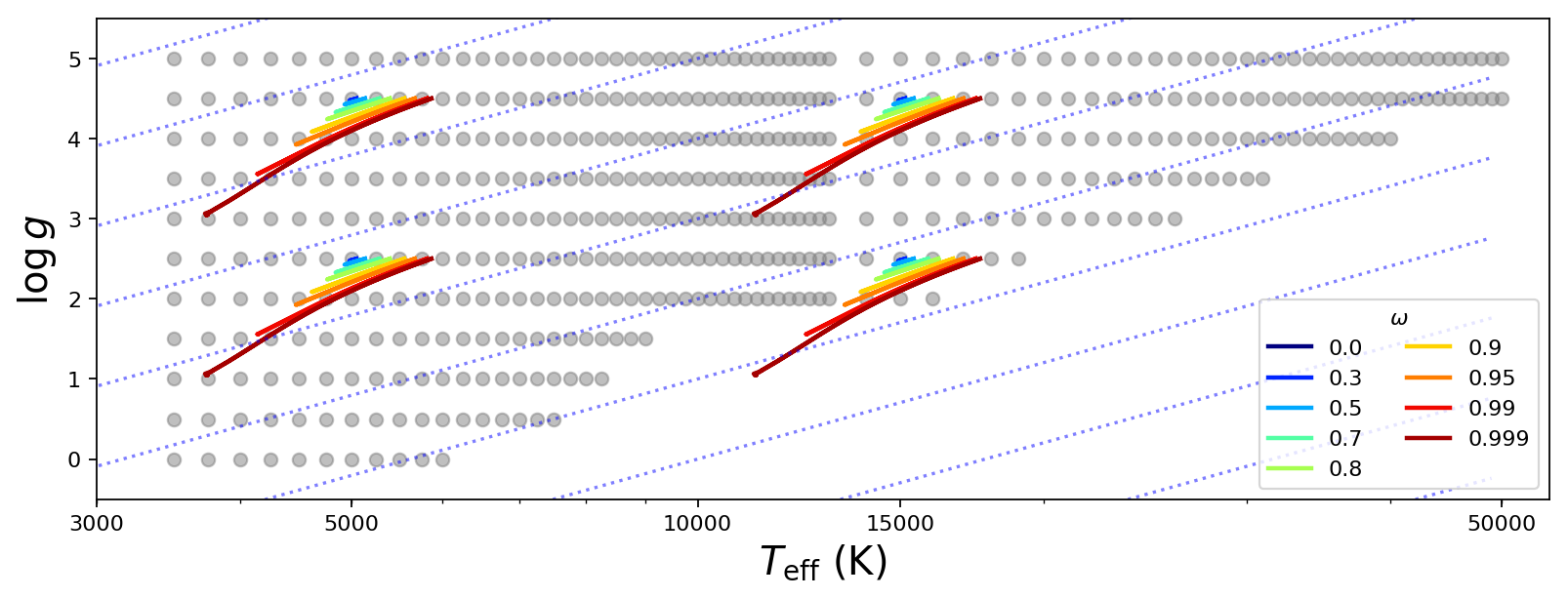}
\caption{\Teff\ and \logg\ values for which $I_\lambda$ is available, from the \citet{castelli04} models of solar metallicity at \url{kurucz.harvard.edu/grids/gridp00/ip00k0.pck19} (gray dots). We superpose the $\Teff$-$\log\geff$ lines spanned by the surface of rotating star models of $\Teff_0=[5000, 15000]$~K and $\log\geffpol=[4.5,2.5]$, at increasing $\omega$ (from blue to red lines, as in the legend). For the sake of comparison, the dotted lines illustrate the slope expected from \citet{vonzeipel24}'s law, $\beta=0.25$, which coincides with the slope found in our models at the limit of small $\omega$. As a consequence of using \citet{espinosalara11} formalism for gravity darkening, our models with high $\omega$ present variable slopes, with mean values approaching $\beta\simeq0.14$ at $\omega>0.99$.}
\label{fig:tefflogg}
\end{figure*}

Finally, the normalization of $F_\lambda^\mathrm{rot}(i)$ can be verified by comparing two quantities: The first is the mean $F_\lambda^\mathrm{rot}(i)$, that is, the average flux seen from random positions uniformly distributed on the sky:
\begin{equation}
  \overline{F_\lambda^\mathrm{rot}(i)} =
       \frac{1}{4\pi}\int_\mathrm{sphere} F_\lambda^\mathrm{rot}(i) \diff \omega' 
       %\nonumber\\
   = \frac{1}{2} \int_0^{\pi/2} F_\lambda^\mathrm{rot}(i) \sin i \diff i
  \label{eq:averageflux}
\end{equation}
(where we use the north-south symmetry of the star to simplify the integral). The second is the average $F_\lambda(\theta)$ over the stellar surface:
\begin{eqnarray}
       \frac{1}{A} \int_\mathrm{surface} F_\lambda(\theta) \diff A = 
       \frac{\sigma\Teff^4}{4\pi R_\mathrm{pol}^2}\, .
\end{eqnarray}
We perform these integrals numerically and find them to be nearly identical, provided that eq.~(\ref{eq:averageflux}) is computed with steps $\Delta i<5^\circ$. The outcoming average spectra are shown with gray lines in Fig.~\ref{fig:spectra}. It is interesting to note that the mean $\overline{F_\lambda^\mathrm{rot}(i)}$ in general presents an excess flux at UV wavelengths, with respect to the reference non-rotating star of same $\Teff_0$, caused by the presence of a hot pole. Nonetheless, the integral of these spectra over $\lambda$ turns out to be the same, and they both respect the Stefan-Boltzmann law for the temperature $\Teff_0$.

%%%%%%%%%%%%%%%%%%%%%%%%%%%%%%%%%%%%%%%%%%%%%%%%%%%%%%%%%%%%%%%%%%%%%%%
\subsection{Limitations}

Our calculations reflect a number of approximations in the modeling of rotating stars. The most fundamental ones consist in using the Roche model to compute the stellar geometry, and the \citet{espinosalara11} approximation to derive the surface distribution of $\Teff$.  A number of works reveal that these are quite good approximations, able to describe both the general behaviour of 2D stellar models, and present observations of rapidly rotating stars \citep{vanbelle12, espinosalara11, claret16}. In particular, the \citet{espinosalara11} model represents a clear improvement over the classical \citet{vonzeipel24} formula, in which $\Teff\propto\geff^\beta$ with $\beta=0.25$. Indeed, \citet{espinosalara11} models produce $\beta$ values variable across the stellar surface, with mean values closer to the expected \citep[and observed; see][]{vanbelle12} $\beta\sim0.14-0.18$ for stars with $\omega$ close to 1. These slopes can be appreciated in the \Teff\ versus \logg\ diagram of Fig.~\ref{fig:tefflogg}. 

Another approximation consists in using plane-parallel model atmospheres, which is quite a reasonable assumption for relatively hot dwarfs. Fortunately, our models are not aimed to describe cool stars with extended atmospheres and convective envelopes, where the validity of these assumptions could be easily questioned.

Apart from the physical assumptions, our calculations are limited by the available libraries of $I_\lambda$, which may not cover the entire possible range of \Teff\ and \logg\ values found in rotating stars.  This is illustrated in Fig.~\ref{fig:tefflogg}, which shows the coverage offered by \citet{castelli04} models. Similar grids are available for metallicity values, [Fe/H], going from $+0.5$ to $-2$ dex. Interpolation inside these grids can provide us with any $I_\lambda$ comprised in the \Teff\ interval from 3500 to 50000~K, and with \logg\ spanning from 5 down to a minimum value which gets larger for higher \Teff. The same plot shows four families of rotating star models at varying $\omega$, for $\Teff_0=[5000, 15000]$~K and $\log\geffpol=[4.5,2.5]$. It is easy to see that models with large $\omega$ may exceed the range of validity of the \citet{castelli04} tables. Two particular cases are worth considering: 
\begin{enumerate}
    \item When regions close to the stellar equator assume $\log\geff$ values smaller than those included in the table, as in the case of the $\Teff_0=15000$~K, $\logg_P=3.5$ modes with $\omega>0.99$. We decide to use the smallest available value of \logg\ in these cases, instead of the correct value, simply because the $I_\lambda$ change slowly with surface gravity. We check this with a more extreme experiment, performed for rapidly rotating stars ($\omega=0.99$) of $\log\geffpol=4.5$ and $3.5$: first we compute the $F_\lambda^\mathrm{rot}(i)$ using the correct calculation of $\log\geff(\theta)$, and then we do the same by setting $\logg$ equal to the polar value everywhere; the resulting $F_\lambda^\mathrm{rot}(i)$ change by just $\lesssim\!5$\% at most. This experiment seems to justify the approach of adopting the $I_\lambda$ tables of smaller $\logg$ values, whenever necessary. In addition, we have to consider that this approximation will only affect hot stars of $\log\geffpol\lesssim3.5$ and $\omega\gtrsim0.9$, which will be very rare (if not absent) in real simulations, since rotation slows down significantly as dwarfs evolve into giants.
    \item Rapidly rotating stars may also exceed the $\Teff$ range of the tables, as it would be the case, for instance, for models with $\Teff_0\lesssim4000$~K and $\omega>0.99$, and those with $\Teff_0$ approaching the upper limit of 50000~K. In these cases, we simply do not compute the $F_\lambda^\mathrm{rot}(i)$. This is probably less of a problem for cool stars, which are observed not to be fast rotators anyway. Rapidly rotating stars with $\Teff_0$ exceeding 30000~K, instead, will not be considered in the present work. 
\end{enumerate}

Another limitation is that we presently do not take into account the broadening of absorption lines by the Doppler effect in rapidly rotating stars. Even if consideration of this effect is relatively simple, the main reasons for this choice are: (1) the computational speed gained by simply adding spectra in bins of $\Teff$, $\log g$, and $\mu'$ -- i.e. without additional bins in radial velocity space; (2) the advantage of producing results independent of the actual stellar radius, i.e., as a function of $\omega$ only; and (3) the fact that we are presently interested in the flux changes expected in {\em broad} filters, generally with widths $\Delta\lambda>500$~\AA, and without sharp edges in the vicinity of strong absorption lines. In comparison, rapidly rotating stars generally have $v\sin i\lesssim400$~km\,s$^{-1}$ \citep[see e.g.][]{royer09} which would translate into a maximum line broadening of $\sim8$~\AA\ in the $V$ band. It is obvious that the Doppler broadening of absorption lines will have to be considered, in addition to our present results, whenever one deals with photometric filters narrower than about $\sim200$~\AA.

%%%%%%%%%%%%%%%%%%%%%%%%%%%%%%%%%%%%%%%%%%%%%%%%%%%%%%%%%%%%%%%%%%%%%%%
\subsection{BC tables for rotating stars}

Setting the equations for $F_\lambda^\mathrm{rot}(i)$ as we have done, we can easily derive bolometric corrections $BC_{S_\lambda}$, that will allow us to transform the bolometric magnitudes, $M_\mathrm{bol}=-2.5\log(L/L_\odot) + M_\mathrm{bol,\odot}$, into absolute magnitudes as a function of $i$, $M_{S_\lambda}= M_\mathrm{bol}- BC_{S_\lambda}$, for any filter transmission curve $S_\lambda$ \citep[cf. eq. (7) in][]{girardi02}:
\begin{eqnarray}
    BC_{S_\lambda} &=& M_\mathrm{bol,\odot} - 2.5\log\left[4\pi(10\,\mathrm{pc})^2 \sigma\Teff_{0}^4/L_\odot\right] \nonumber\\
    &&+ 2.5 \log\left( 
        \frac{\int_{\lambda_1}^{\lambda_2}\lambda F_\lambda^\mathrm{rot}(i) 10^{-0.4A_\lambda} S_\lambda \diff\lambda}{\int_{\lambda_1}^{\lambda_2}\lambda f_\lambda^0 S_\lambda \diff\lambda}
    \right) - m_{S_\lambda}^0 \,.
    \label{eq:bc}
\end{eqnarray}
This equation is generic and could be applied for any set of $F_\lambda^\mathrm{rot}(i)$ and  interstellar extinction curves $A_\lambda$. For the moment, we simplify the analysis by setting $A_\lambda=0$. 

Then, it is a matter of fact that we have already large tabulations of the BCs for non-rotating stars, as a function of $\Teff$, \logg, and [Fe/H] \citep[e.g.][]{girardi02,girardi08}. These grids work quite well in predicting the multi-band photometry of single stars, and some of them exist in different versions -- for instance, with an improved spectral resolution, or with chemical abundances better matching the stars observed in different galactic samples, etc. Therefore, one may find it convenient to work with the ``changes in BC caused by the rotation'', rather than with the absolute BCs themselves. For that, we can simply tabulate the
    \begin{equation}
        \Delta BC_{S_\lambda}(i,\omega, \Teff_0, \log\geffpol) = 2.5\log\left( 
        \frac{\int_{\lambda_1}^{\lambda_2}\lambda F_\lambda^\mathrm{rot}(i,\omega,\Teff_0) S_\lambda \diff\lambda}{\int_{\lambda_1}^{\lambda_2}\lambda F_\lambda(\Teff_0, \log\geffpol) S_\lambda \diff\lambda}
    \right)
    \end{equation}
and apply them to derive the changes in the magnitudes of rotating stars, as a function of $(i,\omega, \Teff_0, \log\geffpol)$. With this approach, we can use the $\Delta BC_{S_\lambda}(i,\omega, \Teff_0, \log\geffpol)$ computed for limited grids -- for instance, including just scaled-solar surface chemical compositions, and just a handful of metallicity values -- for a wide variety of rotating stars. This is the approach we will follow in future applications. 
A suitable set of $\Delta BC_{S_\lambda}$ tables are being inserted in the TRILEGAL code \citep[see][]{girardi05, marigo17} for the production of isochrones and simulated stellar populations in many different photometric systems. They are also part of the YBC database of BCs and interpolating routines (\url{http://stev.oapd.inaf.it/YBC/}; Chen et al., in prep). 

\begin{figure*}
\centering
\includegraphics[width=\textwidth]{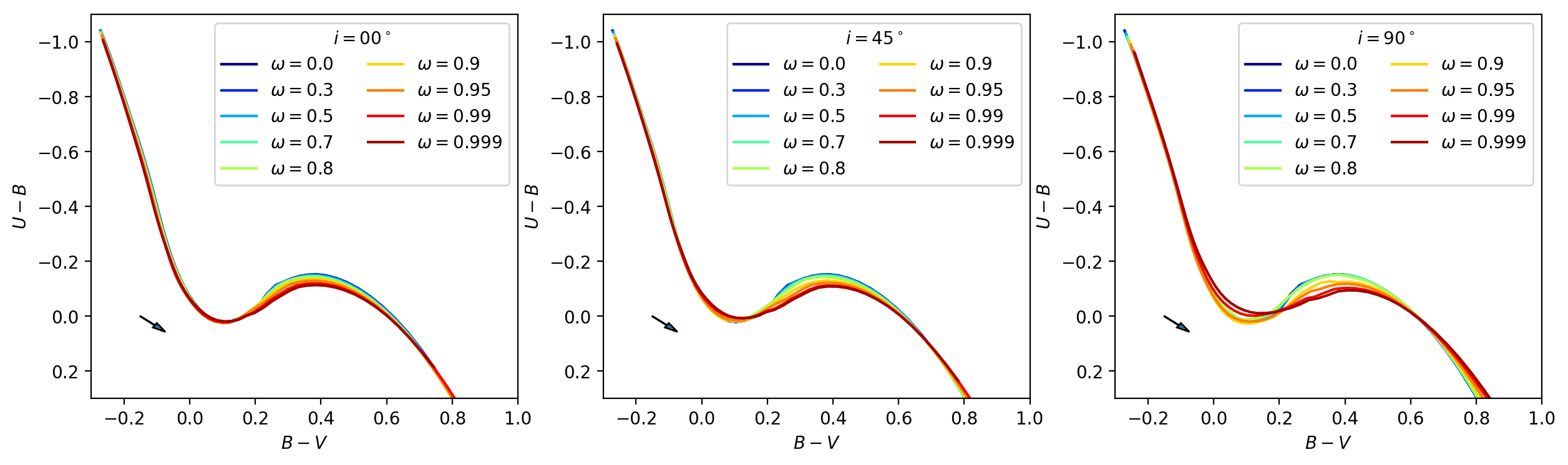}
\caption{The $U-B$ versus $B-V$ diagram for our dwarf models of $\log\geffpol=4.5$ and $\feh=-0.5$~dex (continuous coloured lines), for several values of $\omega$ and for three different inclinations going from the pole-on ($i=0^\circ$) to equator-on ($i=90^\circ$) configurations. 
}
\label{fig:ccd_ourScurve}
\end{figure*}

%%%%%%%%%%%%%%%%%%%%%%%%%%%%%%%%%%%%%%%%%%%%%%%%%%%%%%%%%%%%%%%%%%%%%%%
\section{Results and applications}
\label{results}

%%%%%%%%%%%%%%%%%%%%%%%%%%%%%%%%%%%%%%%%%%%%%%%%%%%%%%%%%%%%%%%%%%%%%%%
\subsection{General behaviour of the spectra and colours}

The $F_\lambda^\mathrm{rot}(i)$ we derive consist essentially in linear combinations of stellar spectra covering a limited range in \Teff, with minor effects coming from the coverage of a limited range in $\logg$. Therefore, they evidently resemble the spectra of non-rotating stars of the same $\Teff_0$, with the presence of either excess flux in the bluer part of the spectrum coming from the hot pole, or ``depleted flux'' caused by the cool equator. For stars in which both the pole and equator are clearly visible, the departures from the spectral shape of non-rotating stars will be the largest. The situation resembles the apparent spectrum of unresolved binaries, which often appear as outliers in colour-colour diagrams, with respect to single stars, due to the different $\Teff$ of the two components.

That said, the colour changes expected for rotating stars are very small compared to those possible for binaries. They become more evident exactly at the $\Teff$ and wavelength ranges in which the spectral features change more rapidly with $\Teff$. One such situation is around the well-known S-shaped curve drawn by single stars in the $U-B$ versus $B-V$ colour-colour diagram (hereafter CCD), which is caused by the appearance of a prominent Balmer jump in main sequence stars with $\Teff\gtrsim7000$~K.
  
Figure~\ref{fig:ccd_ourScurve} illustrates the $U-B$ versus $B-V$ diagram for a subset of our models. As can be seen, fast rotators are observed to ``spread'' aside the S-curve of non-rotating stars, but much more prominently for pole-on stars than for equator-on. This is because, when clearly observed, the hot pole quickly dominates the emitted flux (since $F_\lambda\propto\Teff^4$), making the star to resemble a non-rotating, hotter star. For rapidly rotating stars seen at high inclinations, the S-curve is still very evident, but appears smoother than the classical one. Only models with $\omega\gtrsim 0.9$ appear more than 0.01~mag away from the $\omega=0$ sequence.

\begin{figure*}
\centering
\includegraphics[width=0.66\textwidth]{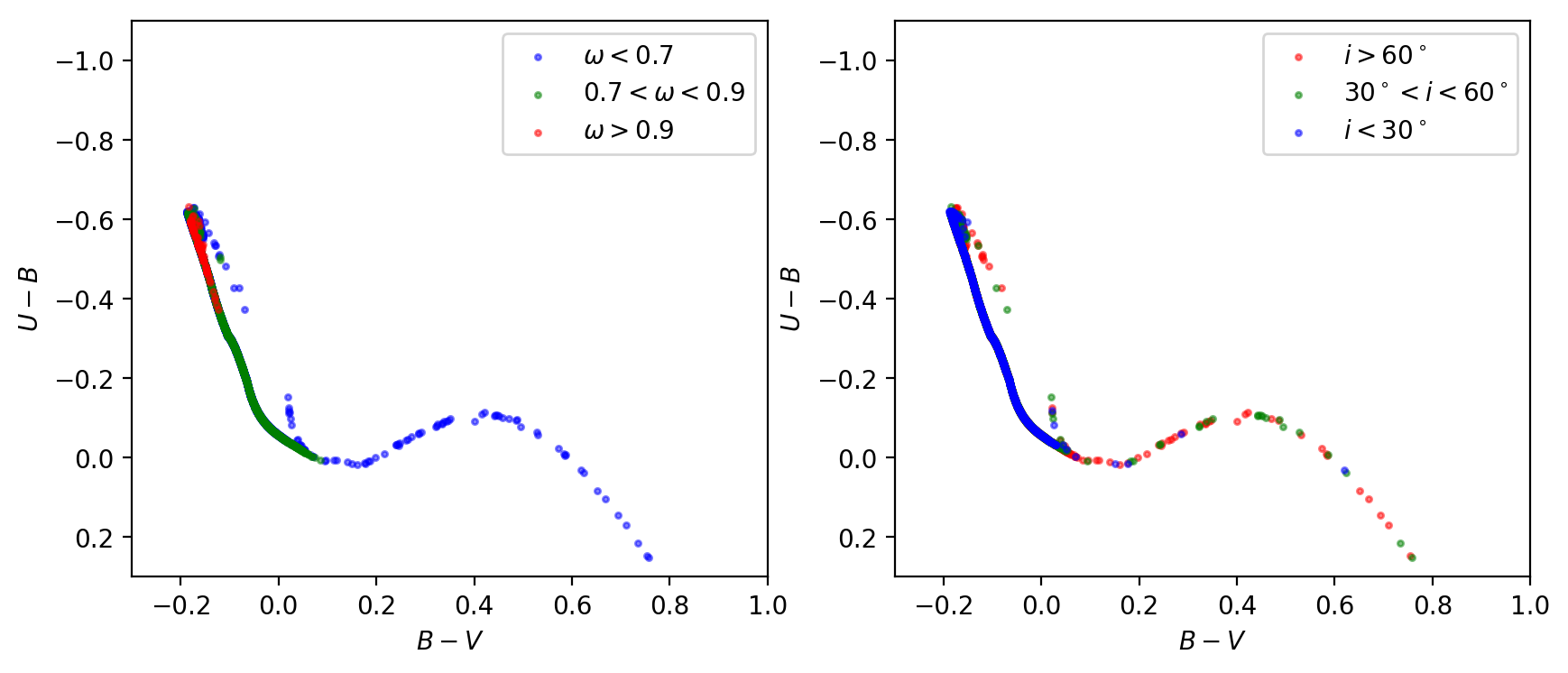}
\caption{The $U-B$ versus $B-V$ diagram for a $Z=0.006$, $10^8$-yr old cluster simulated with the SYCLIST code \citep{georgy14}, containing $10^4$ stars with a uniform distribution of initial $\omega$ observed at random orientations. The left-hand panel colour-code the stars according to their present $\omega$, whereas the right-hand does the same according to $i$.}
\label{fig:ccd_syclist}
\end{figure*}

%%%%%%%%%%%%%%%%%%%%%%%%%%%%%%%%%%%%%%%%%%%%%%%%%%%%%%%%%%%%%%%%%%%%%%%
\subsection{Comparison with other approaches}

Detailed modeling of gravity darkening in rotating stars is not a novelty \citep[see for instance][]{aufdenberg06, lovekin06, espinosalara11, espinosalara13, claret16}, but few are the {\em extended grids} of rotating models aimed at performing population synthesis of such stars, hence considering wide ranges of stellar parameters such as $\Teff_0$, $\log\geffpol$, $\omega$, and $i$, as well as appropriate interpolation and simulation tools. 

The currently most popular set of such tools is SYCLIST from \citet{georgy14}. They follow a different approach to simulate the photometry of rotating stars: They first derive the apparent luminosity and effective temperature of the star at several $i$, which they refer to as $L_\mathrm{MES}$ and $\Teff_\mathrm{MES}$, and then convert these quantities into absolute magnitudes and colours by means of tables of colour-\Teff\ calibration and BC tables. Their procedure includes corrections for the limb darkening, and is consistent from the point of view of the energetics. However, they do not explicitly compute the apparent spectra at several $i$, and use colour-\Teff\ calibration and BC tables that were derived for non-rotating stars. As a consequence, their rotating models cannot present the small deviations from the colour-colour relations of non-rotating stars that ours do. This can be appreciated in Fig.~\ref{fig:ccd_syclist}, which illustrates the $U-B$ versus $B-V$ diagram of a star cluster simulated with SYCLIST and including a significant star-to-star spread in both $\omega$ and $i$, yet describing a very narrow S curve.

This is an important point, since present-day HST photometry of Magellanic Cloud star clusters with broad turn-offs and/or split main sequences, and suspected to contain rapidly rotating stars, include highly-precise multi-band photometry of a few clusters \citep[see e.g.][]{milone16, milone17, goudfrooij18}. The analysis of their colour-colour plots can benefit from considering the expected deviations from the standard colour-colour relations.

Recently, also the MIST project has made available models with rotation and applied them to the detailed modeling of star clusters \citep{gossage18}. From the details provided in \citet{paxton19}, we can conclude that their photometry is computed using essentially the same method as in SYCLIST, that is by first computing projected values of $L$ and $\Teff$ at several $i$, and then applying bolometric corrections as a function of these $\Teff$s. 

\begin{figure}
\centering
\includegraphics[width=\columnwidth]{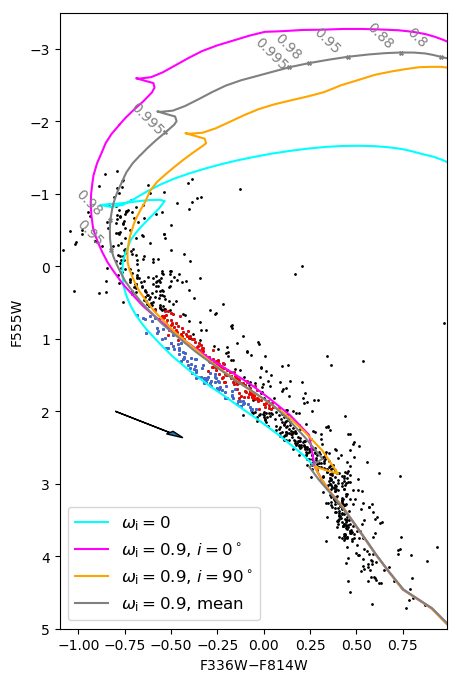}
\caption{HST photometry of the main sequence of NGC~1866, corrected for a true distance modulus of $(m\!-\!M)_0=18.43$~mag and a foreground extinction of $A_V=0.28$~mag \citep[see][]{goudfrooij18}. This particular CMD presents a striking split main sequence, presumably caused by the presence of both slow rotators (in blue) and fast rotators (in red). They correspond to the bMS and rMS in \citet{milone17}, respectively. The continuous lines are PARSEC 250-Myr old isochrones of metalliciy $Z=0.006$, computed with initial values of $\omegai=0$ (cyan) and $\omegai=0.9$ (magenta, orange and gray, for $i=0^\circ$, $i=90^\circ$, and for the mean of stars observed at random orientations, respectively). Stars in the lower main sequence, with masses smaller than 1.8~\Msun, are all computed with $\omegai=0$. The arrow illustrates the reddening vector corresponding to $A_V=0.1$~mag. Dots and labels along the $\omegai=0.9$ mean isochrone are indicating the changes in the actual $\omega$ values as stars move away from their zero age main sequence. 
}
\label{fig:hrd1866}
\end{figure}
 
%%%%%%%%%%%%%%%%%%%%%%%%%%%%%%%%%%%%%%%%%%%%%%%%%%%%%%%%%%%%%%%%%%%%%%%
\subsection{An example: fast rotators in NGC~1866}

As discussed above, present models introduce a new aspect that can be explored in the identification and study of fast rotators, namely that such stars should follow colour-colour relations slightly different from those of non-rotating stars. The effect should be present independently of evolutionary aspects -- at least in less evolved stars in which the initial rotation has not slowed down significantly. 
 
Let us look at one of the best known examples of star clusters containing both fast and slow rotators in their main sequence: the $\sim$\,200-Myr old LMC star cluster NGC~1866. A careful study by \citet{milone17} revealed a clearly double main sequence in the F336W$-$F814W versus F814W CMD of this cluster, in the magnitude range $19 \la \mathrm{F814W} \la 21$. Comparison with SYCLIST models suggested that the blue main sequence (bMS) is caused by slow rotators, while the red sequence (rMS) is caused by stars rotating as fast as $\omega=0.9$. A third, even redder sequence, is also present and likely caused by the presence of approximately equal-mass binaries. Other aspects of these sequences are surprising, as discussed in \citet{milone17}: although the ratio between bMS and rMS varies with magnitude and spatial position in the cluster, nearly 2/3 of the observed stars were found in the rMS, hence suggesting an extremely high fraction of very fast rotators in this cluster.
    
The HST data for this cluster has been reduced independently by us \citep[F336W, F438W, F555W, and F814W passbands; see][]{goudfrooij18}. Fig.~\ref{fig:hrd1866} shows the F555W versus F336W$-$F814W CMD for this cluster. With its long wavelength baseline, the F336W$-$F814W colour allows to spread the main sequence features the most. Only stars with photometric errors smaller than 0.05 mag in all passbands are plotted. For comparison, we overplot PARSEC isochrones \citep[][and work in prep.]{costa19} computed both without rotation ($\omegai=0$), and with fast rotation (i.e. for an initial value of $\omegai=0.9$). The isochrones have been assigned magnitudes directly applying the BCs we derived in this paper (eq.~(\ref{eq:bc})). The $\omegai=0.9$ isochrones are plotted for the two extreme values of $i$ ($0^\circ$ and $90^\circ$), as well as for their mean properties when observed at random orientations. Compared to the models presented in \citet{costa19}, the present ones are calculated with an updated version of the PARSEC code which takes into account the mass loss enhancement due to the rotation, following the prescriptions by \citet{heger00}. The updated code, the new tracks and isochrones and the comparison with other codes, will be presented in a future paper (Costa et al., in prep.). Here, suffice it to recall that the tracks and isochrones include the changes in $\omega$ with respect to its initial value $\omegai$. For stars of intermediate mass, $\omega$ generally increases as the core contracts towards the end of the main sequence, and then sharply decreases as the star evolves into a red giant -- as can be seen in the figure for the $\omegai=0.9$ case. However, in the magnitude interval where the bMS and rMS are well delineated, the value of $\omega$ is essentially identical to $\omegai$.

Essentially, the plot in  Fig.~\ref{fig:hrd1866} reproduces the initial suggestion by \citet{milone17}, that the two components of the cluster split main sequence indicate two very different rotational velocities: a bMS with $\omega=0$ and a rMS with $\omega=0.9$. In the latter case, $i$ values ranging from $0^\circ$ to 90$^\circ$ cause a spread in colour--magnitude space that runs almost parallel to the main sequence, hence it does not cause significant broadening of the isochrones unless in its evolved part, close to and above the turn-off.
   
\begin{figure*}
\centering
\includegraphics[width=0.66\textwidth]{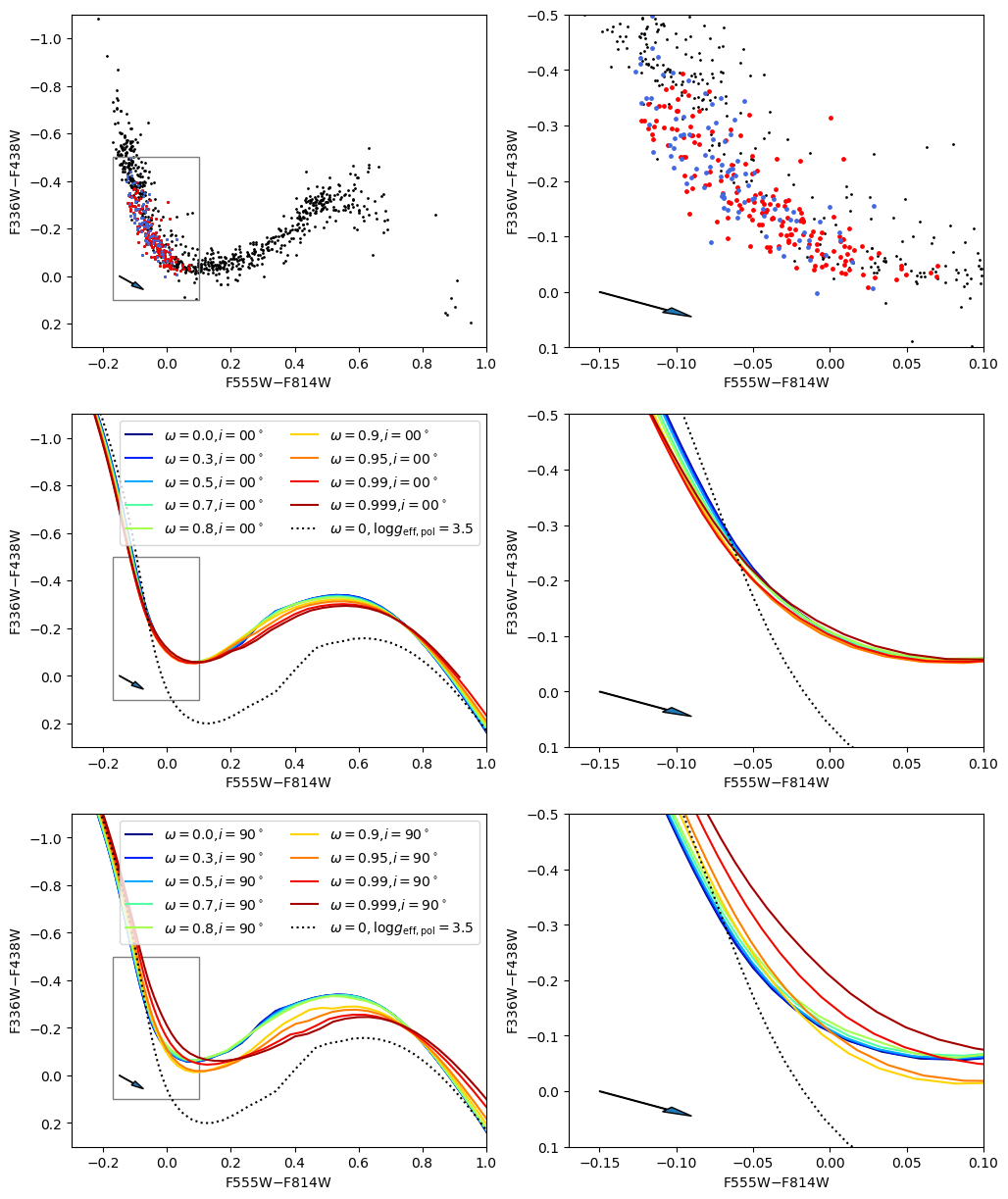}
\caption{The potential effect of rotation in present CCDs from HST data. The top panels show the photometry of NGC~1866 \citep[Fig.~\ref{fig:hrd1866} and][]{goudfrooij18}) corrected by a foreground extinction of $A_V=0.28$~mag, in the F555W$-$F814W versus F336W$-$F438W diagram, around its characteristic S-shaped feature. Again, suspected slow and fast rotators in the cluster main sequence are marked in blue and red, respectively. The right panel zooms in the region occupied by bMS and rMS stars (gray box in left panel), revealing that there is no evident offset between the two groups of stars in the CCD.
The central and bottom rows show the same diagrams as derived from our models for different values of $\omega$: the middle row is for $i=0^\circ$, the bottom row for $i=90^\circ$. All the sequences drawn with continuous lines are for dwarfs with $\log\geffpol=4.5$. To illustrate the dependence on surface gravity, the dotted line shows a single sequence of non-rotation models with $\log\geffpol=3.5$. In all panels, the arrow indicates the reddening vector corresponding to $A_V=0.1$~mag, for comparison.}
\label{fig:ccd1866}
\end{figure*}
  
However, our spectral models tell more about these different main sequences, than can be presumed from the CMD alone: they tell that the fast rotators should present slightly different colour-colour sequences than the slow rotators. Therefore, in Fig.~\ref{fig:ccd1866} we present the F336W$-$F438W versus F555W$-$F814W CCD, derived by the combination of WFC3/UVIS and ACS/WFC photometry. This colour-colour combination captures the best sensitivity we have to changes in the stellar $\Teff$, in the current HST data for NGC~1866: indeed, while F336W$-$F438W should be more sensitive to size of the Balmer jump developing at $\Teff\gtrsim7000$~K, F555W$-$F814W measures the more gradual change in the slope of the red part of the spectra, over a much wider range of $\Teff$. This colour combination also presents the S-shaped sequence characteristic of the classic $U-B$ versus $B-V$ plot, already shown in Fig.~\ref{fig:ccd_ourScurve}.

It turns out that the bMS and rMS merge into a single, indistinguishable sequence in this CCD, as illustrated in the top panels of Fig.~\ref{fig:ccd1866}. The relative mean shift between these two sequences, as measured at intermediate colours (F336W$-$F438W between $-0.3$ and $-0.1$) amounts to 0.01~mag at most. 

Let us then interpret the CCD, according to our models. As illustrated in the middle and bottom panels of Fig.~\ref{fig:ccd1866}, the only models to present significant displacements from the reference stellar locus defined by slow rotators, are those with $\omega>0.95$ and observed nearly equator-on ($i=90^\circ$). Therefore, observing such a narrow CCD sequence means that either the fraction of fast rotators with $\omega>0.95$ is negligible, or that these fast rotators are all observed nearly pole-on. This limit on the fraction of very fast rotators seems in agreement with the \citet{milone17} initial interpretation, which attributed $\omega=0$ for the bMS, and $\omega=0.9$ for the rMS. 

Remarkably, the {\em combined} constraints from the CMD and CCDs of NGC~1866 point to a significant fraction of fast rotators with velocities close to the critical value (more specifically, close to $\omega=0.9$), but at the same time apparently excluding the values immediately close to it (i.e. excluding the interval with $\omega>0.95$, unless for $i\sim0^\circ$). If we assume that fast rotators follow a random distribution of inclination angles, this points to a convergence of fast rotators to a very narrow range of $\omega$, which is intriguing to say the least.
 
It suggests a careful analysis of the data using both evolutionary tracks and population synthesis tools, taking into consideration the additional geometric effects described in this work. These steps will be performed in a subsequent paper (G. Costa et al., in prep).

Although the NGC~1866 observations are particularly intriguing, several other young LMC clusters also show a similarly bifurcated upper MS \citep[see, e.g.,][]{correnti17, milone18}. On the other hand, in the CMD of the NGC~1831 cluster in \citet{goudfrooij18}, the upper MS is far less bimodal in appearance, suggesting a more uniform distribution of $\omega$ at the older age of NGC~1831 ($\sim$\,700 Myr). These aspects will be discussed in more detail in a forthcoming paper (M. Correnti et al., in prep). In all cases, it will be interesting to check whether colour-colour diagrams for these clusters -- whenever available and of sufficient quality -- provide additional constraints to the distributions of rotational velocities and stellar inclinations. 
 
%%%%%%%%%%%%%%%%%%%%%%%%%%%%%%%%%%%%%%%%%%%%%%%%%%%%%%%%%%%%%%%%%%%%%%%
\section{Conclusions}
\label{conclu}

In the present work, we propose a formalism for the inclusion of geometric effects and gravity darkening in models for the synthetic photometry of rotating stars. We compute the outcoming flux and bolometric corrections expected for stars spanning a very large range in mean effective temperature, surface gravity at the pole, metallicity, rotational velocity, and observed inclination. Although the computations are illustrated just for a few filters, they are available for over 50 photometric systems covering all major instrumentation and surveys, hence extending the database of such calculations for non-rotating stars in the PARSEC database of isochrones (see \url{http://stev.oapd.inaf.it/cmd}). The next step is to actually apply these models to a new family of PARSEC evolutionary models with rotation (Costa et al., in prep.), hence closing the loop. 

Our models inevitably rely in some approximations (e.g. the Roche model, the \citealt{espinosalara11} formula, the plane-parallel atmospheres) which can be improved as more realistic grids of rotating models are built. Despite these approximations, the present approach at least opens the way for a systematic consideration of the possible colour-colour effects in population synthesis models, which is still missing in the study of both Magellanic Cloud and Galactic open clusters.  We demonstrate that our models predict small, but non-negligible deviations of the fast rotators from the colour-colour relations of non-rotating stars. Identifying these deviations in the highly-precise HST photometry of star clusters might be perfectly possible, especially for very fast rotators seen nearly equator-on. This was illustrated by comparing our predicted colour-colour relations with those observed in the main sequence of the LMC cluster NGC~1866, which is thought to contain a sequence of rapidly-rotating stars. Our models suggest an upper limit of $\omega=0.95$ to the rotational velocity of this sequence or, alternatively, that any stars rotating faster than such a limit are seen nearly pole-on.

That said, the predicted deviations from the colour-colour relations of non-rotating stars might be identifiable also in other future databases of high-precision photometry, other than the HST one, at least for stars very close to their critical break-up rotation. The basic observational requirement is having photometry with an accuracy of $\sim0.01$~mag in several filters spanning the maximum possible wavelength range -- and possibly including at least one filter blueward of the Balmer jump, as illustrated in this work. The future combination of very large databases accurate to the millimagnitude level, such as Gaia+LSST+Euclid+WFIRST, might provide such an interesting database for looking for very fast rotators in the Milky Way. Of course, such a search might deal with a number of complications, such as the intrinsic colour-colour spread caused by star-to-star variations in metallicity and extinction, and the more subtle problem of distinguishing unresolved binaries from rotating stars. Therefore, searches of these photometric effects should better start in star clusters, to minimize metallicity and extinction variations.

%%%%%%%%%%%%%%%%%%%%%%%%%% 
\section{Acknowledgements}
We acknowledge the support from the  ERC Consolidator Grant funding scheme ({\em project STARKEY}, G.A. n. 615604). Support for this project was provided by NASA through grant
HST-AR-15023 from the Space Telescope Science Institute, which is
operated by the Association of Universities for Research in Astronomy, Inc.,
under NASA contract NAS5-26555.
 
%%%%%%%%%%%%%%%%%%%%%%%%%%%%%%%%%%%%%%%%%%%%%%%%%%%%%%%%%
%to be commented before sending to editor
%\bibliographystyle{mn2e/mn2e_new} %style mn.bst
%\bibliography{references} % your references file.bib
%
%to be uncommented before sending to editor

%
\label{lastpage}
\end{document}